# MDA in Capillary for Whole Genome Amplification

Junji Li [a], Na Lu [a], Xulian Shi [a,b,c], Yi Qiao [a], Liang Chen [a], Mengqin Duan [a], Yong Hou [b,c], Qinyu Ge [a], Yuhan Tao [a], Jing Tu [a,*], Zuhong Lu [a,*]

[a] *State Key Laboratory of Bioelectronics, School of Biological Science and Medical Engineering, Southeast University, Nanjing, 210096, China*
[b] *BGI-Shenzhen, Shenzhen 518083, China*
[c] *China National GeneBank, BGI-Shenzhen, Shenzhen 518120, China*
*\* Correspondence: jtu@seu.edu.cn, zhlu@seu.edu.cn.*

**Abstract:** Whole genome amplification (WGA) plays an important role in sample preparation of low-input templates for high-throughput sequencing. Multiple displacement amplification (MDA), a popular isothermal WGA methods, suffers a major hurdle of highly uneven amplification. Optimizations have been made in the past by separating the reagents into numbers of tiny chambers or droplets in microfluidic devices, which significantly improves the amplification uniformity of MDA. However, skill barrier still exists for biological researchers to handle chip fabrication and droplet manipulation. Here, we present a novel MDA protocol, in-capillary MDA (icMDA), which significantly simplifies the manipulation and improves the uniformity of amplification by dispersing reagents in a long quasi-1D capillary tubing. We demonstrated that icMDA is able to accurately detect SNVs with higher efficiency and sensitivity. Moreover, this straightforward method employs neither customized instruments nor complicated operations, making it a ready-to-use approach for most laboratories.

**Introduction**

Over the past 10 years, whole genome amplification (WGA), which is a bridge from low initial genomic DNA to high-throughput sequencing library (Shapiro et al. 2013; Zhang et al. 1992), has attracted lots of attention. It is generally acknowledged that WGA methods are divided into two categories: polymerase chain reaction (PCR) based methods (Lao et al. 2008; Telenius et al. 1992) and multiple displacement amplification (MDA) based methods (Dean et al. 2002; Dean et al. 2001; Lasken 2009). To guarantee high physical coverage of genome and good uniformity of amplicons, non-specificity amplification is desired, especially in the early stage of amplifying (Chen et al. 2014). Due to the preference of primers, product of PCR methods covers less regions of genome but always performs better in uniformity (Huang et al. 2015). The isothermal methods based on MDA can cover most of the genome, but are challenged by the uneven amplification, which results in bad uniformity within a genome and low resolution of heterozygous information between genomes (Hou et al. 2015). Hybrid methods, such as MALBAC (Lu et al. 2012; Zong et al. 2012) and PicoPLEX, use random priming and poikilothermic pre-amplification in the early stage, followed by PCR amplification with limited circles. Coverage and uniformity are both improved because the poikilothermic pre-amplification generate quasi-linear multiple products, and the following limited PCR cycles induce less bias. Nevertheless, in comparison with pure PCR and isothermal methods, hybrid methods achieve intermediate results (Gawad et al. 2016). Moreover, replacing high fidelity phi29 polymerase by Bst polymerase introduces more bias and complicates the final products (Chen et al. 2014). In spite of less uniformity, MDA is still mostly adopted because of its simple procedure, relatively reliable result and efficient performance.

Microfluidic devices provide powerful and flexible platforms for WGA and extend its application. Due to the isothermal amplification, MDA is more compatible than PCR to be integrated into a microfluidic system and has been widely employed. One of the most significant applications is single cell sequencing (Fritzsch et al. 2012), which has been introduced to the researches of environmental microbial diversity, genetic alterations of cancer cells and germline gene transfer (Gawad et al. 2016; Wang and Navin 2015). However, most of these applications merges MDA with microfluidic system directly without any optimization. The uneven amplification still restricts MDA from being efficiently applied in the cases associated with copy numerical relationships, including single-nucleotide variations (SNVs), copy-number variations (CNVs) and loss of heterozygosity (LOH) (Navin 2014). The development of physical segregation of analyte molecules in microfluidic devices brings the dawn for improving the amplification uniformity of MDA. In earlier works, reagents were directly separated into numbers of nanoliter-scale chambers (Marcy et al. 2007) or wells (Gole et al. 2013) in microfluidic devices. Amplification bias is greatly reduced and more heterozygous genomic diversities are detected. Later on, with the permeating of droplet-based microfluidic devices, there emerged several droplet-MDA protocols (Fu et al. 2015; Nishikawa et al. 2015; Sidore et al. 2016). These techniques further shrink reactors into picoliter, making them markedly effective against contaminations and amplification bias (Tay et al. 2015). The dual effect of increasing parallel-assay numbers and limiting the influencing range of adverse reactions from contaminations and amplification-preferred templates significantly improves the performance and efficiency of these methods. Recently, droplet-based MDA was applied in a metagenome study (Hammond et al. 2016), suggesting that quantitative relationships in metagenomes from low biomass environments can be well recovered. Although the fabrication and operation of microfluidic devices are convenient, to construct and debug a whole droplet reaction system from scattered materials and instruments is a hard work for biological or chemical researchers (Chen et al. 2016). Alternatively, commercial droplet generators and accessories are expensive. Moreover, recovering MDA products from emulsion is complicated and part of the amplicons might be wasted in this step. In this communication, we present a novel MDA method, in-capillary MDA (icMDA), which significantly improves the uniformity of the isothermal amplification. In this method, MDA is performed in slender capillary tubing, and no experimental appliance and accessory needs to be specifically customized. By comparing with conventional MDA, which is usually reacted in a micro-centrifuge tube, we demonstrate that simply conducting conventional MDA reaction in capillary tubing with small inner diameter efficiently suppresses amplification bias.

**Materials and methods**

In our experiment, all of the glass and plastic consumables were properly cleaned, dried and UV-treated before use. Purified genomic DNA of YH-1 cells (immortalized cells of a Han Chinese individual, provided by BGI Shenzhen) was extracted (QIAamp DNA Mini Kit, Qiagen), then quantitated using commercially available kits (Qubit dsDNA HS Assay Kit, ThermoFisher Scientific). Commercialized MDA kit (REPLI-g MIDI kit, Qiagen) was used according to the manufacturer's protocol, with few modifications. We added 20 ng genomic DNA into the reaction buffer for a total 100 microliter volume. After denaturing and neutralizing, the reaction buffer containing genomic DNA was divided into two parts, 50 microliters for conventional in-tube MDA and the rest 50 microliters for icMDA. DNA polymerase was added separately. Afterwards, conventional MDA tube was incubated in a thermo cycler at 30°C for 16 h, followed by 3 min inactivation at 65°C. The rest division of reaction mixture was injected into a polytetrafluoroethylene (PTFE) capillary tubing (320 micron inner diameter, Adtech Polymer Engineering Ltd.) of 1 meter length in which the 50 microliters reaction system formed a liquid column of 0.62 meter length. This capillary was pre-treated with 0.1% w/w bull serum albumin (BSA) for 1 h. Then we sealed both ends of the capillary by epoxy glue (5 Minute Epoxy, Devcon) and coiled this capillary around a metal cord-winder. All of the preparations were performed at 4°C to avoid the amplification starting in advance. Eventually, the capillary-coiling winder was immersed in a 30°C water bath for 16 h, then moved into another water bath of 65°C for 5 min (Fig. 1). Amplification products in centrifuge tube and capillary tubing were collected, purified and constructed into Illumina libraries respectively. Each library was sequenced on an Illumina HiSeq 4000 instrument using $2 \times 150$ paired-end reads. Qualified sequencing data was aligned to the hg19 reference genome using the BWA short sequence alignment software (Li and Durbin 2009), and then statistically analyzed through QualiMap 2 (Okonechnikov et al. 2016). SNVs were called by using SAMtools (Li et al. 2009) and Varscan 2 (Koboldt et al. 2012). Further results were calculated and exhibited by Perl scripts.

**Results and discussion**

icMDA reaction produced comparable products in comparison with conventional in-tube assay (S1, Supporting information) in the same amount of time, which demonstrated that icMDA is as efficient as in-tube MDA. We inferred that a certain degree of geometry alteration of reaction space from cube-like to linear does not affect the global aspects of reaction process. In addition, almost no extra expense and hand-on time is required for executing icMDA. By substituting standard PTFE capillary tubing with customized microfluidic devices, broader geometry adjusting range could be achieved with a little extra expense and limited time-and-labor consumption.

After analyzing the sequencing data, we found that icMDA reads covered wider areas in genome than in-tube MDA reads with same sequencing depth (S2, Supporting information). This advantage becomes more evident when comparing the coverage of genome with high sequencing depth. Then we checked genomic areas of at least $20 \times$ sequencing depth, icMDA reads covered 13.28% more areas. The results supported that smaller quantities of icMDA products were able to recover the target genome. By $13 \times$ sequencing depth, products of icMDA recovered over 95% of the genome, and in contrast, product of in-tube MDA required $25 \times$ sequencing effort (S3, Supporting information) to achieve the same performance. To evaluate the genomic coverage distribution, we separated the whole genome of hg19 into bins of fixed 40 Kb size, then calculated the standardized average depth of each bin (detailed in S4, Supporting information). Fixed-size binning was used because it is able to show more primitive results than those optimized binning strategies, such as variable binning. The read depth of each bin was normalized by corresponding results of unamplified bulk sample to exclude the sample specificity. The genomic coverage distribution of icMDA reads was significantly even than that of in-tube MDA reads in the whole genome (S5, Supporting information). The coefficient of variation (CV) of icMDA and in-tube MDA at this bin size were 0.37 and 0.78 respectively. From the histogram of the read depth over the entire chromosome X (Fig. 2a), we observed that all the bins of the icMDA sequencing data were generally even in depth with CV of 0.37, while numerous abnormal bins were observed in the sequencing data of in-tube MDA (CV = 0.83). Moreover, if we zoomed in and narrowed the bin size to 2 Kb, the uniformity superiority was still prominent (Fig. 2b). We also plotted Lorenz curves to validate coverage uniformity of genome (Fig. 3a). Central diagonal line represents theoretical perfect uniformity of read distribution. By comparing the Lorenz curves of bulk, icMDA and in-tube MDA samples, we found that icMDA sample showed closer performance to the unamplified bulk sample, while the conventional MDA sample deviated obviously from the bulk and icMDA samples. All above curves were obtained from ~$30 \times$ mean sequencing depth data. There is no difference of amplification conditions between icMDA and in-tube MDA except the reactors used. Therefore, the geometry change of the reactor seems to be the major cause of the uniformity improvement of icMDA.

In order to quantitatively compare the influence of reactors with different geometries, we introduced a simulation of average distance between the central unit and the other units, which is relative to the change of inner diameter of capillary tubing (S6, Supporting information). In this simulation, reactors in different geometries were divided into small reaction units according to the template concentration. The central unit of any cylindrical reactors is tend to be mostly influenced by other units. From the simulation result, we observed significantly sharp increase of average distance as the inner diameter of the reactor decreases (Fig. 3b). Here we use inner diameter rather than internal surface area, which is a more appropriate parameter for irregular-shape reactor, to simplify calculation and construction of diffusion model.   Increased average distance reveals that most reaction units in the capillary are remarkably isolated with a certain proof-of-concept unit, thus the local depletion of big molecules (Welch et al. 1995), such as primers, are not able to be replenished timely and abundantly by the units far apart while the few units nearby only provide limited replenishment. In this study, 50 microliters reaction system formed a liquid column of 0.62 meter length in the thin capillary with the average distance of 0.17659 meter, which is about 100 times of the average distance in a cylinder whose diameter equals its length to approximate the experimental condition in tube. Therefore, most of the rest units are more distant with the central unit in capillary than in tube. As a result, over-amplification (de Bourcy et al. 2014) caused by the snowballing effect of random priming would be confined in a local range by using the quasi-1D capillary tubing, thus ensuring the high overall amplification uniformity.

In terms of SNV detection, icMDA gives better detection rate for both homozygous and heterozygous SNVs on chromosome 1 of diploid YH-1 cell (Table 1). We identified 83.42% high-confidence SNVs using icMDA, in contrast to 67.40% by conventional MDA with the same data size. We then examined the allelic dropout (ADO) rate by counting the loss-of-heterozygosity events from the comparison of SNVs in icMDA/MDA and in bulk. Due to the relatively high initial quantity, the ADO rate of icMDA is low to 3.12%, better than 3.35% of MDA. Improvement of both detection efficiency and ADO rate results from the better amplifying uniformity of icMDA. Next, we counted the false positives and errors by comparing SNVs in icMDA/MDA with SNVs in bulk. Both false positive rate and error rate of icMDA is in the same magnitude but superior to those of in-tube MDA. This fact indicates that geometric change of reactor in icMDA does not affect the accuracy derived from phi29 polymerase. Subsequently, we reduced the input data to $10 \times$, icMDA still maintains relatively higher detection rate for SNVs, meaning that icMDA is more sensitive to discover SNVs with low sequencing depth than conventional MDA (S7, Supporting information).

By using icMDA, amplification uniformity is highly improved without any decrease of amplifying efficiency. Almost no extra instrument but few commercially available accessories are required in icMDA experiment. The isolation of reaction units can also be accomplished by using micro-reactor MDA (Gole et al. 2013; Marcy et al. 2007) or emulsion MDA (Fu et al. 2015; Nishikawa et al. 2015; Sidore et al. 2016), and higher amplification uniformity are also obtained by these approaches. However, either micro-reactor MDA or emulsion MDA requires relatively precision instruments, delicate operations and considerable additional cost (Table 2). The simple experimental procedures and operations make icMDA more appropriate for trace samples as simple procedure causes less unexpected loss of sample. Besides, abandon of emulsions not only simplifies some experience-required process like microfluidic device commissioning, monodisperse droplet generation and demulsification, but also avoids the probable interference from surfactants and heterogeneous system. Instead of physical isolation, icMDA maintains the connectivity of the whole reaction system but increases the difficulty of interaction between reactants by spreading them over a long physical distance. This feature preserves the stability and precision of amplification because diffusion could alleviate the drastic change in concentrations of ions and small molecules during the reaction process.

**Conclusions**

We proposed an improved MDA technique, icMDA, which conducts conventional MDA reaction in a quasi-1D capillary tubing. By comparing with conventional in-tube MDA product with the same initial reaction system, we demonstrated that our approach dramatically enhances the amplifying uniformity throughout the entire genome by suppressing the over-amplification in a local range. In terms of SNV detection, icMDA exhibits higher efficiency and sensitivity for both homozygous and heterozygous SNVs while maintaining the same high-level accuracy as conventional MDA. Different from previously improved methods like micro-reactor MDA and emulsion MDA, our method requires little additional cost, no customized equipment or accessories, and is simple and convenient in experimental operations. icMDA presents important improvements in amplification uniformity, experiment cost and operation complexity. We believe that this novel method could provide a more convenient approach to access the genomic information, especially for the studies requiring precise quantitative relation in a genome or among genomes.

**Acknowledgements**
This work was supported by the National Key Project of China (No. 2016YFA0501600), project 61571121 of National Natural Science Foundation of China and the Fundamental Research Funds for the Central Universities.

**Appendix A. Supporting information**
Supplementary data associated with this article can be found in the online version at:

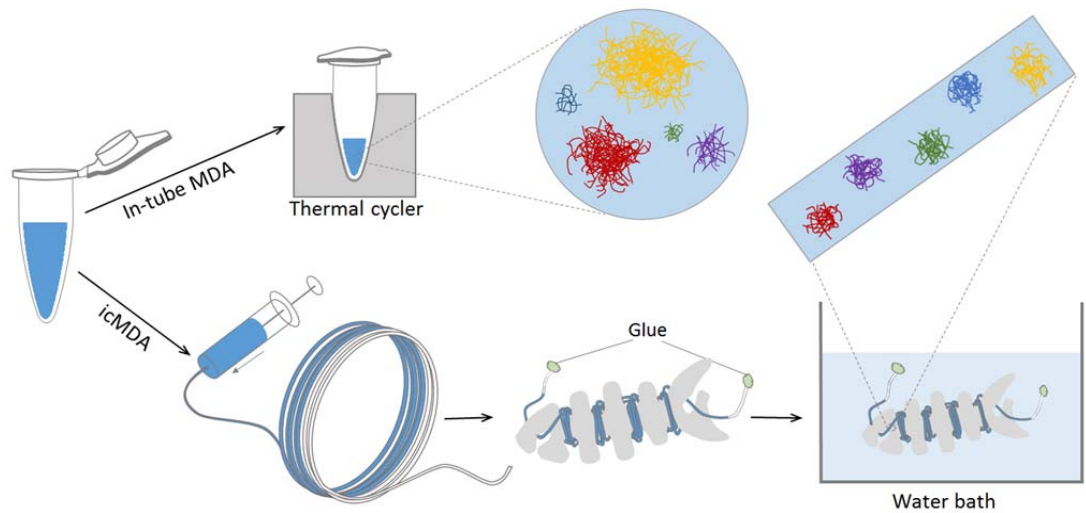

**Fig. 1.** Schematic illustration of icMDA experimental process. Purified genomic DNA of YH-1 cell line was added into MDA reagents, then partitioned in two for conventional in-tube MDA and icMDA. The icMDA portion was then injected into PTFE capillary tubing which was wrapped around a metal cord-winder. After sealed the both ends, the capillary was placed in water bath for isothermal incubation at 30°C for 16 hours. In conventional in-tube MDA, reagents were distributed in three-dimensional space, leading preferred templates been over-amplified by timely reactant supplement. In icMDA, reaction units were uniformly distributed in a narrow capillary. The quasi-1-dimentional geometry strongly restricted diffusion of large molecules such as enzymes, templates and primers, thus cut off the positive feedback of over-amplification, resulting in an even amplification.

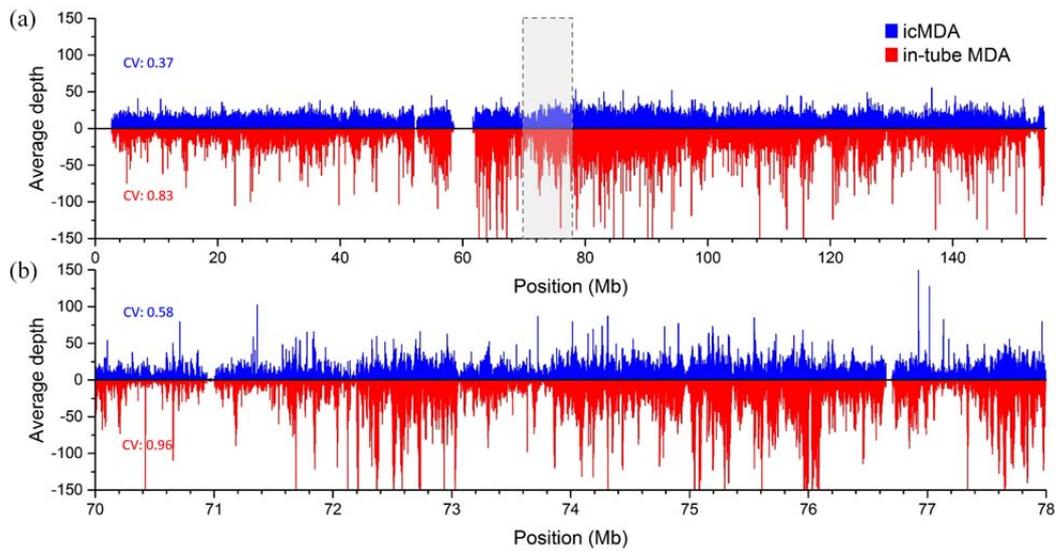

**Fig. 2.** Comparison of sequencing coverage between icMDA reads and in-tube MDA reads. (a) Read depth distribution over chromosome X of YH-1 cell line. The entire chromosome was divided into 40 Kb bins, and we calculated the standardized mean coverage depth of icMDA reads (blue bars, above X axis) and in-tube MDA reads (red bars, below X axis) in each bin. (b) Zoom in on the 70 Mb to 78 Mb region of chromosome X (dashed box with gray background in (a)). Bin size was adjusted to 2 Kb to ensure the same sampling rate with (a).

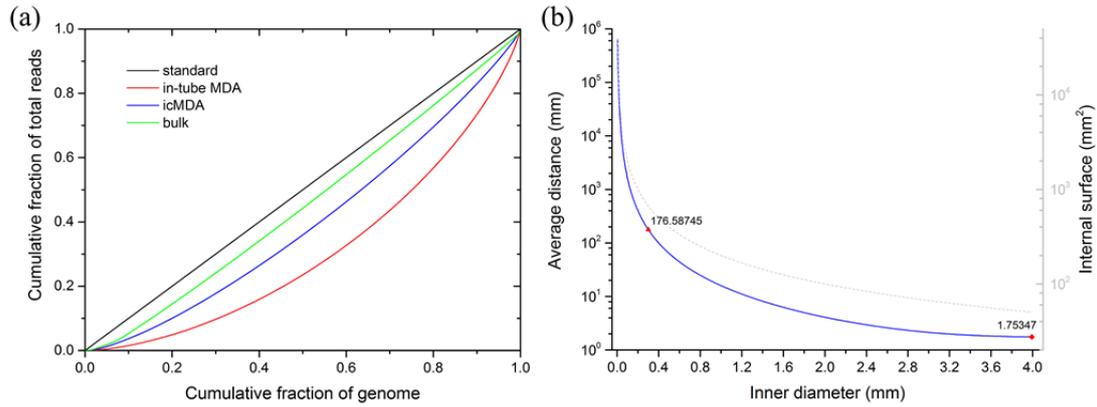

**Fig. 3.** Analysis of read distribution uniformity. (a) Lorenz curves of bulk, icMDA and in-tube MDA samples, in which the cumulative fraction of total reads were plotted as a function of the cumulative fraction of genome, depict read coverage uniformity across whole genome. The diagonal line represents perfectly uniform coverage, as is the reference line of curves calculated from reads of unamplified bulk sample (green), icMDA products (blue) and conventional in-tube MDA products (red). All samples were sequenced at ~30× depth. (b) Simulation of the average distance between central reaction unit to other units. Reaction space of different geometry was divided into minimized units of which the number equals to the number of templates. Red triangle and rhombus on the curve respectively represents situations of conducting MDA in a capillary of 300 micron inner diameter and in a diameter-equal-to-height cylinder, which has the similar geometry to conventional reaction tube. The gray dash shows internal surface area of capillary reactors with different inner diameters.

**Table 1.** Summary of the comparison between different methods for SNV detection on chromosome 1 of normal diploid YH-1 cell

| Parameter | Sample type* | | |
|---|---|---|---|
|  | icMDA | MDA | Bulk |
| Total SNVs | 236464 | 191056 | 283476 |
| Detection rate | 83.42% | 67.40% | N/A |
| Heterozygous SNVs | 99254 | 79123 | 117367 |
| Detection rate | 84.57% | 67.42% | N/A |
| Homozygous SNVs | 137210 | 111933 | 166109 |
| Detection rate | 82.60% | 67.39% | N/A |
| ADO rate | 3.12% | 3.35% | N/A |
| SNV error rate | 0.05% | 0.13% | N/A |
| False-positive rate | 7.50% | 8.13% | N/A |

* Calculation was based on sequencing data of larger than 30 × data size.

**Table 2.** Comparison of icMDA with reported MDA methods.

| | MDA | | Micro-reactor MDA** | | Emulsion MDA | | icMDA | |
|---|---|---|---|---|---|---|---|---|
| **Amplifying performance** | | | | | | | | |
| Uniformity | low | | medium | | high | | high | |
| Efficiency | high | | medium | | high | | high | |
| **Additional setup costs*** | | | | | | | | |
| Equipment | N/A | | ~$0-8000 | | ~$1000 | | N/A | |
| Mould | N/A | | ~$100-2000 | | ~$100 | | N/A | |
| **Additional process costs and hands-on times*** | time | cost | time | cost | time | cost | time | cost |
| Device fabrication | N/A | N/A | 3h-7h | ~$10-50 | 3h | $20 | N/A | N/A |
| System adjusting | N/A | N/A | 0-1h | N/A | 30min | N/A | N/A | N/A |
| Experimental process | N/A | N/A | 5min-1h | N/A | 25min | $1 | 15min | $0.5 |

\* Costs are estimated according to the lowest quoted prices in China then converted into dollars. As lithography instrument is not commonly equipped in biological laboratories, microfluidic-involved methods take the unified processing mode of outsourced fabrication of moulds and on-site fabrication of microfluidic devices. N/A represents not applicable.

\*\* Differences between micro-reactor MDA methods are large. Typically, High degree of automation requires more complex and expensive chip fabrication and control but significantly reduces operating time.

# Supplementary Information

**S1. Amplification efficacies of conventional in-tube MDA and icMDA.**

The reaction volume of both assays was 50μl, reaction time was 16 hours, and inner diameter of icMDA capillary was 320 micron. Products were fully vortexes then diluted to 0.01x and quantitated by Qubit 2.0.

|         | In-tube MDA (1x, ng/μl) | icMDA (1x, ng/μl) |
|---------|-------------------------|-------------------|
| Assay 1 | 413                     | 416               |
| Assay 2 | 357                     | 348               |
| Assay 3 | 366                     | 403               |
| Average | 379                     | 389               |

**S2. Whole genome sequence analysis of conventional in-tube MDA and icMDA products.**

|                          | In-tube MDA | icMDA   |
|--------------------------|-------------|---------|
| Data size (GB)           | 123.24      | 128.32  |
| Mapped reads (MB)        | 754.86      | 757.99  |
| Average depth ($\times$) | 35.25       | 34.71   |
| Coverage (at least 1$\times$) | 91.15% | 92.18%  |
| Coverage at least 5$\times$   | 83.92% | 90.37%  |
| Coverage at least 10$\times$  | 73.64% | 85.61%  |
| Coverage at least 20$\times$  | 54.24% | 67.52%  |

**S3. Genome recovery from raw sequence reads.**

Smaller amount of icMDA reads can recover more target genome than conventional in-tube MDA reads.

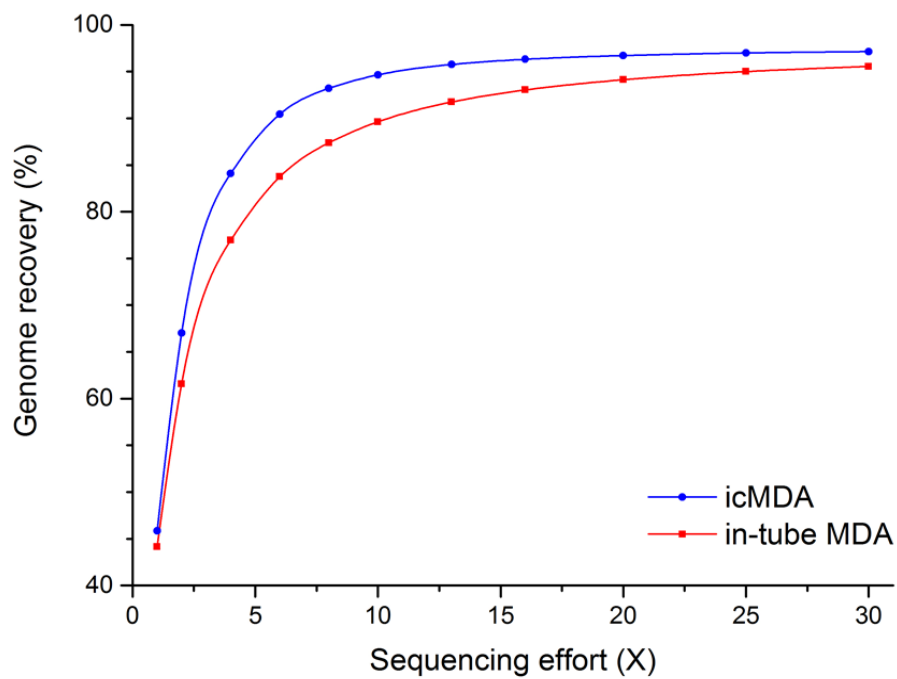

**S4. Fixed-size binning and standardization of mean depth in each bin.**

Several regions, such as highly repetitive regions and regions around the centromeres, always show aberrant read depth. This aberrance can be fixed by dividing uniquely mapped reads into variable-length intervals across the genome. In this work, we used fixed-size binning combined with standardization by unamplified bulk data, thus transforming mean depth of each bins from absolute quantities into relative quantities independent with alignment and genome structure. To obtain the standardized coefficient, we divided every single-base counts by the mean count of each chromosome. For icMDA and in-tube MDA data, we divided their every single-base counts by the coefficients of each corresponding position, then assigned them into 40Kb bins and calculated the mean depth of each bin. This standardized fixed-size binning efficiently reduces over- and under-expression of highly repetitive regions while retaining data changes caused by over-amplification.

**S5. Comparison of sequencing coverage between icMDA reads and in-tube MDA reads.** Chromosomes of YH-1 cell line was divided into 40 Kb bins respectively, and we calculated the standardized mean coverage depth of icMDA reads (blue, above X axis) and in-tube MDA reads (red, below X axis) in each bin.

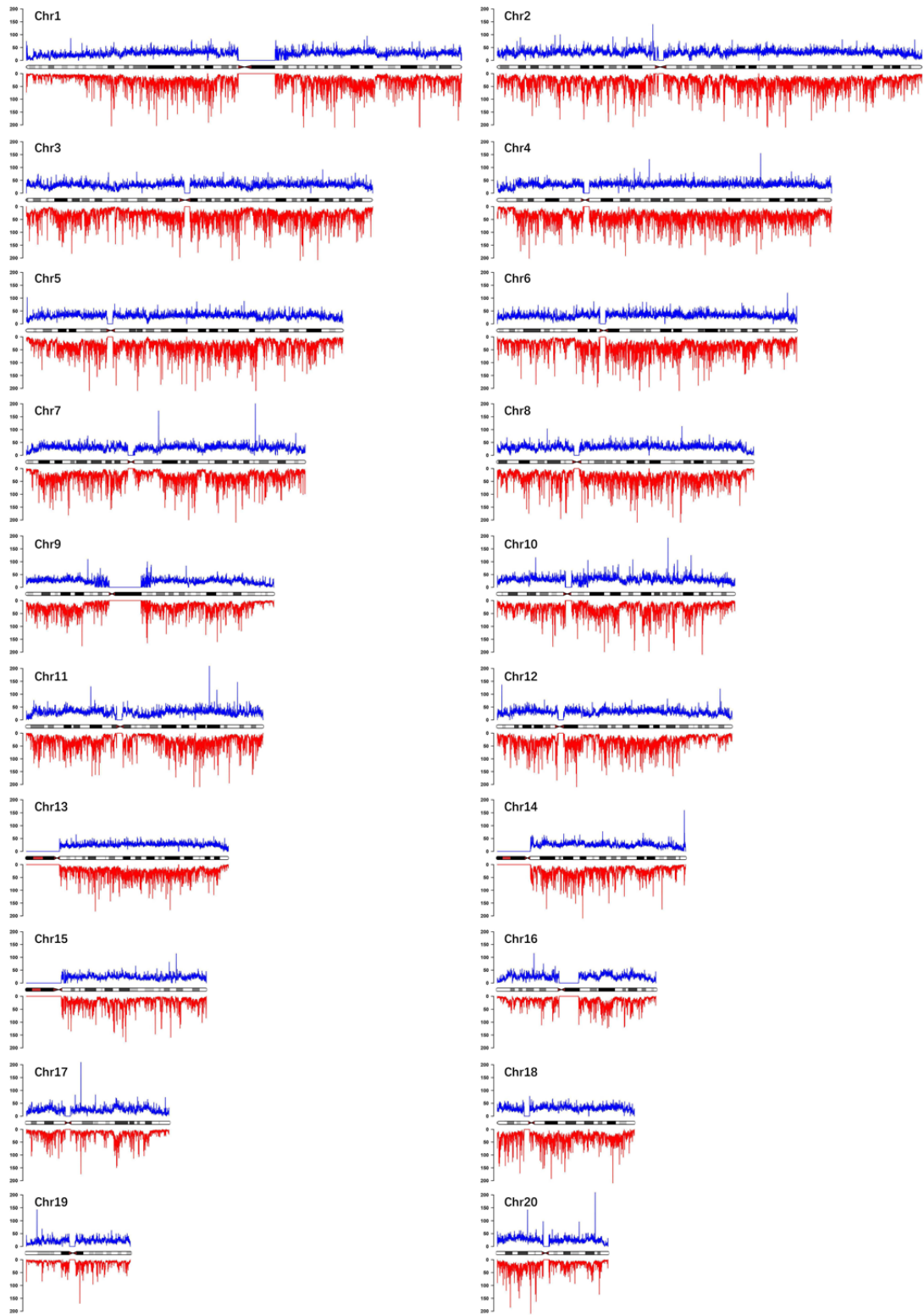

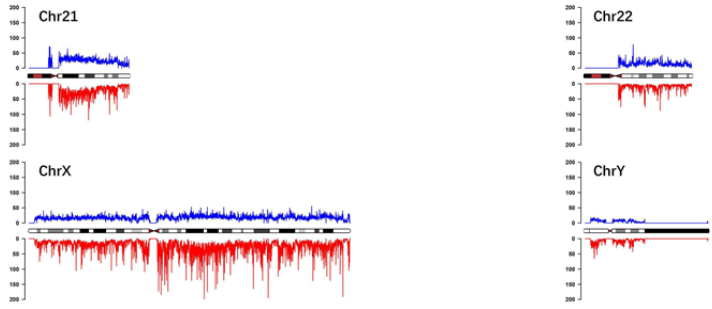

**S6 Simulation of average central distance of minimized reaction units.**

Both capillary tubing and conventional reaction tube was equivalent to approximate cylinders with different lengths and bottom diameters. Hence, the length of reaction space can be achieved by equation: $l=V/(\pi * r^2)$, where $V$ is a constant volume of 50 microliters, $l$ and $r$ represents length and inner diameter of the capillary tubing, respectively. The maximum value of $r$ is 3.992 millimeters, corresponding to the experimental conditions in tube of which the geometry was abstracted into a cylinder whose diameter is equal to its length. Minimum of $r$ is 1.084 microns, calculated from formula $r_{min} = \sqrt[3]{\dfrac{\overline{l_t} \cdot \overline{M_r}}{2\pi c N_A}}$, where $\overline{l_t}$ is the mean length of template DNA, $\overline{M_r}$ is the mean relative molecular mass per base pairs, $c$ is the concentration of template DNA and $N_A$ is Avogadro constant. Physical significance of $r_{min}$ is the distance between any two nearest templates, when $r$ is in the range of $r_{min}$ to $r_{max}$, the aforementioned distance remains a constant correlative with template concentration. We averagely sampled 1000 points between $r_{min}$ to $r_{max}$, then calculated each average central distance $d_{ave}$. central distance is an simulated parameter, meaning the distance between an arbitrary location of template and the location of the central template in the reaction space. Calculation of $d_{ave}$ is accelerated by downsampling $r$ and $l$ with the parameters of the minimum between $r/r_{min}$ and 1000, and the minimum between $l/r_{min}$ and 1000, respectively.

**S7 Comparison of SNV-detection efficiency between icMDA and conventional MDA using 10× sequencing data.**

|  | Heterozygous SNVs | Homozygous SNVs | Total SNVs |
|---|---|---|---|
| **Bulk (30× data)** | | | |
| SNVs | 117367 | 166109 | 283476 |
| **Bulk (10× data)** | | | |
| SNVs | 109328 | 160243 | 269571 |
| Detection rate | 93.15% | 96.47% | 95.09% |
| **icMDA (10× data)** | | | |
| SNVs | 84514 | 153878 | 238392 |
| Detection rate | 72.01% | 92.64% | 84.10% |
| **MDA (10× data)** | | | |
| SNVs | 71441 | 143701 | 215142 |
| Detection rate | 60.87% | 86.51% | 75.89% |